\documentclass[12pt]{article}
\usepackage{oldgerm}
\usepackage{yfonts}
\usepackage{euler}
\usepackage{graphics}
\usepackage{bm}
\usepackage{graphicx}
\usepackage{epstopdf}
\usepackage{amsmath}
\usepackage{amssymb}
\usepackage{amstext}
\usepackage{amscd}
\usepackage{amsfonts}
\usepackage{appendix}
\usepackage{color}
\DeclareMathAlphabet{\EuFrak}{U}{euf}{m}{n}
\DeclareMathAlphabet{\EuScript}{U}{eus}{m}{n}

\newcommand{\dq}[1]{\frac{\partial {#1}}{\partial q}}
\newcommand{\dt}[1]{\frac{\partial{#1}}{\partial t}}
\newcommand{\dts}[1]{\frac{\partial^2{#1}}{\partial t^2}}
\newcommand{\dx}[1]{\frac{\partial{#1}}{\partial x}}
\newcommand{\dxs}[1]{\frac{\partial^2{#1}}{\partial x^2}}

\newcommand{\corch}{\left[1+(1-q)z\right]}
\newcommand{\limq}[1]{\lim\limits_{q\rightarrow1}{#1}}
\newcommand{\val}[1]{\left.{#1}\right|_{q=1}}
\newcommand{\z}{\frac{i}{\hbar}(px-Et)}

\newcommand{\corchf}{\left[1+\frac{i}{\hbar}\frac{(1-q)}{q}Et\right]}
\newcommand{\corchg}{\left[1+\frac{i}{\hbar}\frac{(1-q)}{\sqrt{2(q+1)}}px\right]}

\newcommand{\nd}{\noindent}

\title{{\bf Perturbative treatment of the  non-linear q-Schr\"{o}dinger
and q-Klein-Gordon equations}}

\author{{D. J. Zamora$^1$, M. C. Rocca$^1$, A. Plastino$^1$,
and G. L. Ferri$^2$}\\
\small{$^1$ La Plata National University and   Argentina's National Research Council}\\
\small{(IFLP-CCT-CONICET)-C. C. 727, 1900 La Plata - Argentina}\\
$^2$ \small{Faculty of Exact and Natural Sciences, La Pampa National University} \\
\small{Uruguay 151, Santa Rosa, La Pampa, Argentina}}

\date{\today}

\begin{document}

\maketitle

\begin{abstract}
\noindent Interesting nonlinear generalization of both Schr\"{o}dinger's and 
Klein-Gordon's equations have been recently advanced by Tsallis, Rego-Monteiro, and Tsallis (NRT) in
 [Phys. Rev. Lett. {\bf 106}, 140601
(2011)]. There is much current activity going on in this area. The non-linearity is governed by a real parameter 
$q$. It is a fact that the ensuing non linear 
q-Schr\"{o}dinger and  q-Klein-Gordon
equations  are natural manifestations of very high energy phenomena,
  as verified by LHC-experiments. This happens for $q-$values close to unity 
	[Nucl. Phys. A {\bf 955}, 16 (2016), Nucl. Phys. A {\bf 948}, 19 (2016)].  It is also 
	well known that q-exponential  behavior is found in
quite different settings. An explanation for such phenomenon was
given in  [Physica A {\bf 388}, 601 (2009)] with reference to
empirical scenarios in which data are collected via set-ups that
effect a normalization plus data's pre-processing. Precisely, the
ensuing normalized output was there shown to be q-exponentially
distributed if the input  data display elliptical symmetry,
generalization of spherical symmetry, a frequent situation.  This
makes it difficult, for q-values close to unity, to ascertain
whether one is dealing with solutions to the ordinary
Schr\"{o}dinger equation (whose free particle solutions are
exponentials, and for which $q=1$) or with its NRT nonlinear
q-generalizations, whose free particle solutions are q-exponentials. In this
work we provide a careful analysis of the $q \sim 1$ instance via
a perturbative analysis of the NRT equations.

\nd {\bf Keywords:} Non-linear Schr\"{o}dinger equation;
Non-linear Klein-Gordon equation; First order solution.

\end{abstract}

\newpage

\renewcommand{\theequation}{\arabic{section}.\arabic{equation}}

\setcounter{equation}{0}

\section{Introduction}

\setcounter{equation}{0}

Empirical data indicates  that power-law behavior in the observed
probability  distribution of interesting  quantities is quite
frequent in the natural world \cite{boccara}. It was  shown in
  \cite{alf} that one reason
 for this phenomenon is detector-normalization. In most  measurement devices
  one finds  a  pre-processing step that impedes that  the
device  be overwhelmed by data of too large amplitude that might
damage he hardware. One then appeals to
 statistically normalization of input data. The data are first centered by
subtraction of their estimated mean, and then scaled  with their
estimated standard deviation. It is shown in \cite{alf} that the
concomitant proceedings transform Gaussian input data into
q-Gaussian output one. We remind that a q-exponential is defined
as

\begin{equation}
\label{eq6.3} e_{q}(x)=[1+(1-q) x]^{\frac{1}{1-q}}; \,\,
\,()e_{q=1}(x) = \exp{x}).\end{equation} In view of the above
empirical considerations, it becomes clear that in the immediate
neighborhood of $q=1$ it is very difficult to ascertain whether
one is dealing with exponentials or with q-exponentials. The first
instance corresponds to free-particle solutions of the celebrated
Schr\"odinger  equation, while the second instance corresponds to
free particle solutions to its q-nonlinear generalization advanced
in \cite{tp3} (see also \cite{tp3bis,tp3bis1}), the so called NRT equation. If we confront a
particle flow, which of the two equations is governing it, the
linear or the nonlinear one?

In order to help finding an adequate answer we carefully study
here perturbative NRT solutions in a $q \sim 1$ scenario. We
hope that our considerations will shed some light on these
matters.

\subsection{Motivation}

The motivation of this paper resides in the fact that both the
q-Schr\"{o}dinger equation and the q-Klein-Gordon
equation are natural manifestations of very high energy phenomena
 \cite{n1,n1b}, as verified by LHC-experiments  \cite{n2}, for which  $q$ is close to unity. 
In such a case,  the two equations mentioned above approach the customary Schr\"{o}dinger
and Klein-Gordon equations, becoming identical to them in the limit
$q\rightarrow 1$. The  q-Schr\"{o}dinger equation is

\begin{equation}
\label{eqm.1}
i\hbar\frac {\partial} {\partial t}\left[
\frac {\psi(\vec{x},t)} {\psi(0,0)}\right]^q+
\frac {\hbar^2} {2m}\triangle
\left[\frac {\psi(\vec{x},t)} {\psi(0,0)}\right]=0,
\end{equation}
and for $q$ close to unity we are allowed to write

\begin{equation}
\label{eqm.2}
\left[\frac {\psi(\vec{x},t)} {\psi(0,0)}\right]^q=
\frac {\psi(\vec{x},t)} {\psi(0,0)}+
(q-1)\frac {\psi(\vec{x},t)} {\psi(0,0)}
\ln\left[\frac {\psi(\vec{x},t)} {\psi(0,0)}\right].
\end{equation}
The second term on the right side of (\ref{eqm.2}) is negligible so that one writes

\begin{equation}
\label{eqm.3}
i\hbar\frac {\partial} {\partial t}\left[
\frac {\psi(\vec{x},t)} {\psi(0,0)}\right]+
\frac {\hbar^2} {2m}\triangle
\left[\frac {\psi(\vec{x},t)} {\psi(0,0)}\right]=0,
\end{equation}
or
\begin{equation}
\label{eqm.4}
i\hbar\frac {\partial} {\partial t}\psi(\vec{x},t)+
\frac {\hbar^2} {2m}\triangle\psi(\vec{x},t)=0,
\end{equation}
the linear, conventional de Schr\"{o}dinger equation.
\vskip 3mm

\nd An analogous situation arises for the  q-Klein-Gordon equation. We have

\begin{equation}
\label{eqm.5}
\square\left[\frac {\phi(\vec{x},t)} {\phi(0,0)}\right]+
\frac {qm^2c^2} {\hbar^2}
\left[\frac {\phi(\vec{x},t)} {\phi(0,0)}\right]^{2q-1}=0.
\end{equation}
For $q$ close to unity one has
\begin{equation}
\label{eqm.6}
\left[\frac {\phi(\vec{x},t)} {\phi(0,0)}\right]^{2q-1}=
\frac {\phi(\vec{x},t)} {\phi(0,0)}+
2(q-1)\frac {\phi(\vec{x},t)} {\phi(0,0)}
\ln\left[\frac {\phi(\vec{x},t)} {\phi(0,0)}\right].
\end{equation}
Once more, the second term on the right is negligible. As a consequence, we can write

\begin{equation}
\label{eqm.7}
\square\left[\frac {\phi(\vec{x},t)} {\phi(0,0)}\right]+
\frac {qm^2c^2} {\hbar^2}
\left[\frac {\phi(\vec{x},t)} {\phi(0,0)}\right]=0,
\end{equation}
or

\begin{equation}
\label{eqm.8}
\square\phi(\vec{x},t)+
\frac {qm^2c^2} {\hbar^2}
{\phi(\vec{x},t)}=0,
\end{equation}
and then, since $q$ is close to unity,

\begin{equation}
\label{eqm.9}
\square\phi(\vec{x},t)+
\frac {m^2c^2} {\hbar^2}
{\phi(\vec{x},t)}=0.
\end{equation}
We see then that it is very important to obtain approximate  q-solutions for the two above scenarios, since these are two instances
 that correspond  to intermediate energies
\cite{n3}.

\setcounter{equation}{0}

\section{First order expansion  of the q exponential
as a solution of the  non-linear NRT q-Schr\"{o}dinger equation}

As a first task we will prove that the first order Taylor expansion,
around $q=1$, of q-exponential function $e_{q}$, is a solution of to
the non-linear q-Schr\"{o}dinger equation advanced in \cite{tp3}
\begin{equation}
\label{eq1.1}
i\hbar\dt{ }\left[\frac{\psi(x,t)}{\psi(0,0)}\right]^q=
H_{0}\left[\frac{\psi(x,t)}{\psi(0,0)}\right]
\end{equation}
$\psi(0,0)$ a fixed wave function, where
\begin{equation}
\label{eq1.2}
H_{0}=\frac{p^2}{2m}=\frac{-\hbar^2}{2m}\nabla^2
\end{equation}
In our case: $\psi\sim e_{q}$,  defined as
\begin{equation}
\label{eq1.3}
e_{q}=[1+\frac{i}{\hbar}(1-q)(px-Et)]^{\frac{1}{1-q}}
\end{equation}
Calling $z=\z$ we deal with
\begin{equation}
\label{eq1.4}
e_{q}=\corch^{\frac{1}{1-q}}
\end{equation}
The function $e_{q}$ tends to the usual exponential when
$q \to 1$
\begin{equation}
e_{q}(z)|_{q\rightarrow 1}=e^z
\label{eq1.5}
\end{equation}
We observe that $\psi(0,0)=1$. Thus, equation (\ref{eq1.1}) reduces to
\begin{equation}
i\hbar\dt{\psi^q}=H_{0}\psi(x,t).
\label{eq1.6}
\end{equation}
\subsection{First order expansion of $\psi =e_{q}$}

We obtain, after a somewhat lengthy manipulation (see Appendix A)
\begin{equation*}
\psi \simeq e^{z}+(q-1)\frac{z^2}{2}e^{z},
\end{equation*}
or
\begin{equation}
\label{eq1.10}
\psi \simeq e^{z}\left[1+(q-1)\frac{z^2}{2}\right].
\end{equation}
Note that the last relation differs form the pure exponential, for q close to unity, just  by the term  $(q-1)x^2/2$ above.\vskip 3mm \noindent
Moreover, we have
\begin{equation}
\label{eq1.11}
\psi \simeq e^{\z}\left[1+(1-q)\frac{(px-Et)^2}
{2\hbar^2}\right]
\end{equation}

We need now the expansion for the second derivative of $e_q$  with
respect to $x$. This involves again some extensive manipulation (see
Appendix A) and one finds
\begin{multline}
\dxs{\psi}=-\frac{p^2}{\hbar^2}e^{\z}\left[q+\frac{2i}{\hbar}(q-1)(px-Et)-\frac{(q-1)}{2\hbar^2}(px-Et)^2\right].
\label{eq1.13}
\end{multline}

Obviously, one also needs the first order expansion of $\psi^q$.
This expansion is in the variable q, around $q=1$, and has no obvious
quantum mechanics counterpart. One writes
\begin{equation*}
\psi^q=e^{q\ln\psi}.
\end{equation*}
Thus,
\begin{equation}
\label{eq1.14}
\dq{\psi^q}=\dq{(q\ln\psi)}\psi^{q}=\left(\ln\psi+\frac{q}{\psi}\dq{\psi}\right)\psi^{q}=\psi^{q}\ln\psi+q\psi^{q-1}\dq{\psi}.
\end{equation}
For q=1 we obtain
\begin{equation*}
\val{\dq{\psi^q}}=\val{\psi}\val{(\ln\psi)}+\val{\dq{\psi}}.
\end{equation*}
Since  we know that
\begin{itemize}
\item $\val{\psi}=e^{z}$
\item $\val{(\ln\psi)}=z$
\item $\val{\dq{\psi}}=\frac{z^2}{2}e^{z},$
\end{itemize}
then,
\begin{equation}
\label{eq1.15}
\val{\dq{\psi^q}}=z e^{z}+\frac{z^2}{2}e^{z},
\end{equation}
\begin{equation}
\label{eq1.16}
\val{\dq{\psi^q}}=\z e^{\z}-\frac{(px-Et)^2}{2\hbar^2}e^{\z}.
\end{equation}
The first order expansion of $\psi^q$ is then, up to a normalization constant
\begin{equation*}
\psi^q\simeq \val{\psi^q}+(q-1)\val{\dq{\psi^q}}.
\end{equation*}
Replacing here equations (\ref{eq1.5}) and (\ref{eq1.16}) we have
\begin{equation}
\label{eq1.17}
\psi^q\simeq e^{\z}+(q-1)\left[\z e^{\z}-\frac{(px-Et)^2}{2\hbar^2}e^{\z}\right].
\end{equation}

Finally, we require the time derivative. The first order time-derivative of $\psi^q$ is
\[\dt{\psi^q}=\frac{-iE}{\hbar}e^{\z}+(q-1)\left[\frac{-iE}{\hbar}e^{\z}+\frac{2E}{\hbar^2}(px-Et)e^{\z}\right.\]
\[\left.+\frac{iE}{2\hbar^3}(px-Et)^2e^{\z}\right]\],
or
\begin{multline}
\label{eq1.18}
\dt{\psi^q}=\frac{-iE}{\hbar}e^{\z}\left[q+\frac{2i(q-1)}{\hbar}(px-Et)-\frac{(q-1)}{2\hbar^2}(px-Et)^2\right].
\end{multline}


\subsection{Solution to the non-linear q-Schr\"{o}dinger  equation}

Replacing equations (\ref{eq1.13}) and (\ref{eq1.18}) into the
non-linear q-Schr\"{o}dinger equation (\ref{eq1.6}), we find that the first order Taylor's expansion  of a q-exponential
is indeed  a solution of this equation with the usual eigenvalue corresponding
to the free particle, $E=p^2/2m$. We have thus achieved self consistency,  which is the first result of the present communication.

\subsection{Comparison between the exact and approximate solutions}

In this subsection we intend making comparisons between
the approximate and exact solutions of q-Schr\"{o}dinger
equation. To this end, we first try to evaluate the modulus of
the ratio of the functions (\ref{eq1.11}) and (\ref{eq1.3}).
(We call to this ratio R).
As an example, we display four figures corresponding to
an electron and a proton with 1 MeV of energy at $t=0$
(Figures 1,2,3, and 4).
Note that, for a range of $x$ very large in terms of atomic or nuclear relevant distances, the ratio is essentially unity. Our approximation can then be deemed very good.

\setcounter{equation}{0}

\section{Separation of variables}

We will see here that, for $\psi(x,t)= f(t)g(x)$, the equations for,
respectively, f and g, keep the form of the nonlinear equation for
$\psi$  (\ref{eq1.1}).
From (\ref{eq1.1}),
the non-linear q-Schr\"{o}dinger equations for separate variables [$\psi(t,x)=f(t)g(x)$]
are given by \cite{tp1}
\begin{equation}
\label{eq2.1}
i\hbar\dt{f^q}=\lambda f(t),
\end{equation}
\begin{equation}
\label{eq2.2}
\frac{-\hbar^2}{2m}\dxs{g}=\lambda g^q.
\end{equation}

We should now prove that the first order expansion  around
$q=1$ of the functions
\begin{equation}
\label{eq2.3}
f(t)=\left[1+
\frac{i}{\hbar}\frac{(1-q)}{q}Et\right]^{\frac{1}{q-1}},
\end{equation}
\begin{equation}
\label{eq2.4}
g(x)=\left[1+
\frac{i}{\hbar}
\frac{(1-q)}{\sqrt{2(q+1)}}px\right]^{\frac{2}{1-q}},
\end{equation}
are solutions of their respective differential equations. This is
done in Supplementary Materials (SM) B.

\subsection{Solution to the differential equations for $f$ and $g$}

From the Appendix B we obtain the two relations

\begin{equation}
\label{eq2.8B} f(t)\simeq
e^{\frac{-iEt}{\hbar}}\left[1+(q-1)\left(\frac{iEt}{\hbar}+\frac{E^{2}t^2}{2\hbar^2}\right)\right],
\end{equation}
and

\begin{equation}
\label{eq2.12B}
\dt{f^q}=-\frac{iE}{\hbar}e^{\frac{-iEt}{\hbar}}
\left[1+(q-1)\frac{E^{2}t^2}{2\hbar^2}+(q-1)\frac{iEt}{\hbar}\right].
\end{equation}

 1) Now,  replacing equations (\ref{eq2.8B}) and (\ref{eq2.12B}) (see
Appendix B) into (\ref{eq2.1}), we observe the first order Taylor
expansion of $f$ is a solution of (\ref{eq2.1}), with
$\lambda=E=p^2/2m$. Again,  self-consistency has been reached. We
pass next to SM C to deal with $g(x)$.

2) From Appendix C we get
\begin{equation}
\label{eq2.22C} g^q\simeq
e^{\frac{ipx}{\hbar}}\left[1+\frac{3(q-1)}{4}
\frac{ipx}{\hbar}-\frac{(q-1)}{4}\frac{p^{2}x^2}{\hbar^2}\right],
\end{equation}
and

\begin{equation}
\label{eq2.18C} \dxs{g}=-\frac{p^2}{\hbar^2}e^{\frac{ipx}{\hbar}}
\left[1-\frac{3(1-q)}{4}\frac{ipx}{\hbar}+\frac{(1-q)}{4}\frac{p^{2}x^2}{\hbar^2}\right].
\end{equation}

 Replacing (\ref{eq2.22C}) and (\ref{eq2.18C}) of Appendix C into
(\ref{eq2.2}), we find that the first order Taylor's expansion is a
solution of this last equation with $\lambda=E=p^2/2m$.

\setcounter{equation}{0}

\section{First order treatment of a q-Gaussian}

As mentioned in the Introduction, this issue is of paramount importance.
From \cite{tp4}, selecting $mq\alpha=1$ in order to
simplify the calculations, we have for the q-Gaussian:
\begin{equation}
\label{eq5.1}
\psi(x,t)=\left\{1+(q-1)\left[a(t)x^2+b(t)x+c(t)\right]\right\}^{
\frac {1} {1-q}}
\end{equation}
where
\begin{equation}
\label{eq5.2}
a(t)=\frac {mq} {1+i\hbar(q+1)t}
\end{equation}
\begin{equation}
\label{eq5.3}
b(t)=\frac {1} {\beta[1+i\hbar(q+1)t]}
\end{equation}
\[c(t)=\left(\frac {1} {q-1}-\frac {1} {4mq\beta^2}\right)
\left[1+i\hbar(q+1)t\right]^{\frac {q-1} {q+1}}+\]
\begin{equation}
\label{eq5.4}
\frac {1} {4mq\beta^2[1+i\hbar(q+1)t]}+
\frac {1} {1-q}
\end{equation}
Writing up to first order

\[a(t)=a_1(t)+(q-1)a_2(t)\]
\[b(t)=b_1(t)+(q-1)b_2(t)\]
\begin{equation}
\label{eq5.5}
c(t)=c_1(t)+(q-1)c_2(t)
\end{equation}
we obtain from (\ref{eq5.2}), (\ref{eq5.3}), and (\ref{eq5.4}):
\begin{equation}
\label{eq5.6}
a_1(t)=\frac {m} {1+2i\hbar t}
\end{equation}
\begin{equation}
\label{eq5.7}
a_2(t)=\frac {m(1+i\hbar t)} {(1+2i\hbar t)^2}
\end{equation}
\begin{equation}
\label{eq5.8}
b_1(t)=\frac {1} {\beta(1+2i\hbar t)}
\end{equation}
\begin{equation}
\label{eq5.9}
b_2(t)=-\frac {i\hbar t} {\beta(1+2i\hbar t)^2}
\end{equation}
\begin{equation}
\label{eq5.10}
c_1(t)=\frac {1} {2}\ln(1+2i\hbar t)-
\frac {i\hbar t} {2m\beta^2(1+2i\hbar t)}
\end{equation}
\[c_2(t)=
\frac {1} {4m\beta^2}+
\frac {i\hbar t} {2(1+2i\hbar t)}-
\frac {(1+3i\hbar t)} {4m\beta^2(1+2i\hbar t)^2}+\]
\begin{equation}
\label{eq5.11}
\frac {1} {8}\ln^2(1+2i\hbar t)-
\frac {1+2m\beta^2} {8m\beta^2}\ln(1+2i\hbar t)
\end{equation}
The first order approximation for the q-Gaussian is now
\[\psi(x,t)=\left\{1-(q-1)\left\{a_2(t)x^2+b_2(t)x+c_2(t)-
\frac {1} {2} \left[a_1(t)x^2+b_1(t)xc_1(t)\right]^2\right\}\right\}\otimes\]
\begin{equation}
\label{eq5.12}
e^{-[a_1(t)x^2+b_1(t)x+c_1(t)]}
\end{equation}
By construction $\psi(x,t)$, as given by (\ref{eq5.11}), is a
 first order solution to (\ref{eq1.1}). Fig. 5 displays the ratio between (\ref{eq5.11}) and 
 (\ref{eq1.1}) versus distance $x$ (in absolute units) for $1-q=10^{-3}$. Remark that the ratio is essentially unity for distances very much larger than atomic or nuclear ones. The approximation is quite good then.

\subsection{Comparison between the exact and approximate solutions}

To have an idea about the quality of the first order
treatment of the qGaussian ans the exact solution we evaluate the 
modulus of the ratio of eq. (\ref{eq5.12}) and (\ref{eq5.1})
in a semi logarithmic scale. This is given in Fig, 5.

\setcounter{equation}{0}
\section{Non-linear q-Klein-Gordon equation}

We should prove that the development of $e_{q}$ is
a solution of the following equation:
\begin{equation}
\label{eq3.1}
\frac{1}{c^2}\dts{F}-\dxs{F}+\frac{qm^{2}c^2}{\hbar^2}F^{2q-1}=0
\end{equation}
This equation was advanced in \cite{tp3} and re-obtained
in \cite{tp2}.

In our case $F=e_{q}$, which, let us remind the reader,  is defined as
\begin{equation}
\label{eq3.2}
e_{q}=[1+i(1-q)(kx-\omega t)]^{\frac{1}{1-q}}
\end{equation}
By analogy to equations (\ref{eq1.11}) and (\ref{eq1.13}) we know the expansions of $e_{q}$ and of its  derivative with respect to $x$, respectively cast as

\begin{equation}
\label{eq4.3}
F \simeq e^{i(kx-\omega t)}\left[1+(1-q)\frac{(kx-\omega t)^2}
{2}\right],
\end{equation}

\begin{multline}
\dxs{F}=-k^{2}e^{i(kx-\omega t)}\left[q+2i(q-1)(kx-\omega t)-\frac{(q-1)}{2}(kx-\omega t)^2\right].
\label{eq4.4}
\end{multline}
We should now calculate the second derivative with respect to $t$
and the first order expansion of $qF^{2q-1}$, and so on. This is
done in Appendix D.

\subsection{Solution  to the  Klein-Gordon equation}
Replacing (\ref{eq3.11}), (\ref{eq3.5}), and (\ref{eq4.4})
into (\ref{eq3.1}) we find that the first order Taylor's expansion
 of the  q-exponential is a solution of this last equation. Again, self-consistency has been achieved.

\setcounter{equation}{0}

\section{Conclusions}

\noindent We have exhaustively analyzed a first order perturbation-treatment (in $q$)  of both the nonlinear q-Schr\"odinger and q-Klein Gordon partial differential equations. We have shown that, for small values of $q-1$, the approximation is quite good. This is of physical relevance because, as  discussed in \cite{n1,n1b}, these q values are the relevant ones in the range of energies of interest for intermediate and high energy physics.

\newpage

\begin{figure}[h]
\begin{center}
\includegraphics[scale=0.6,angle=0]{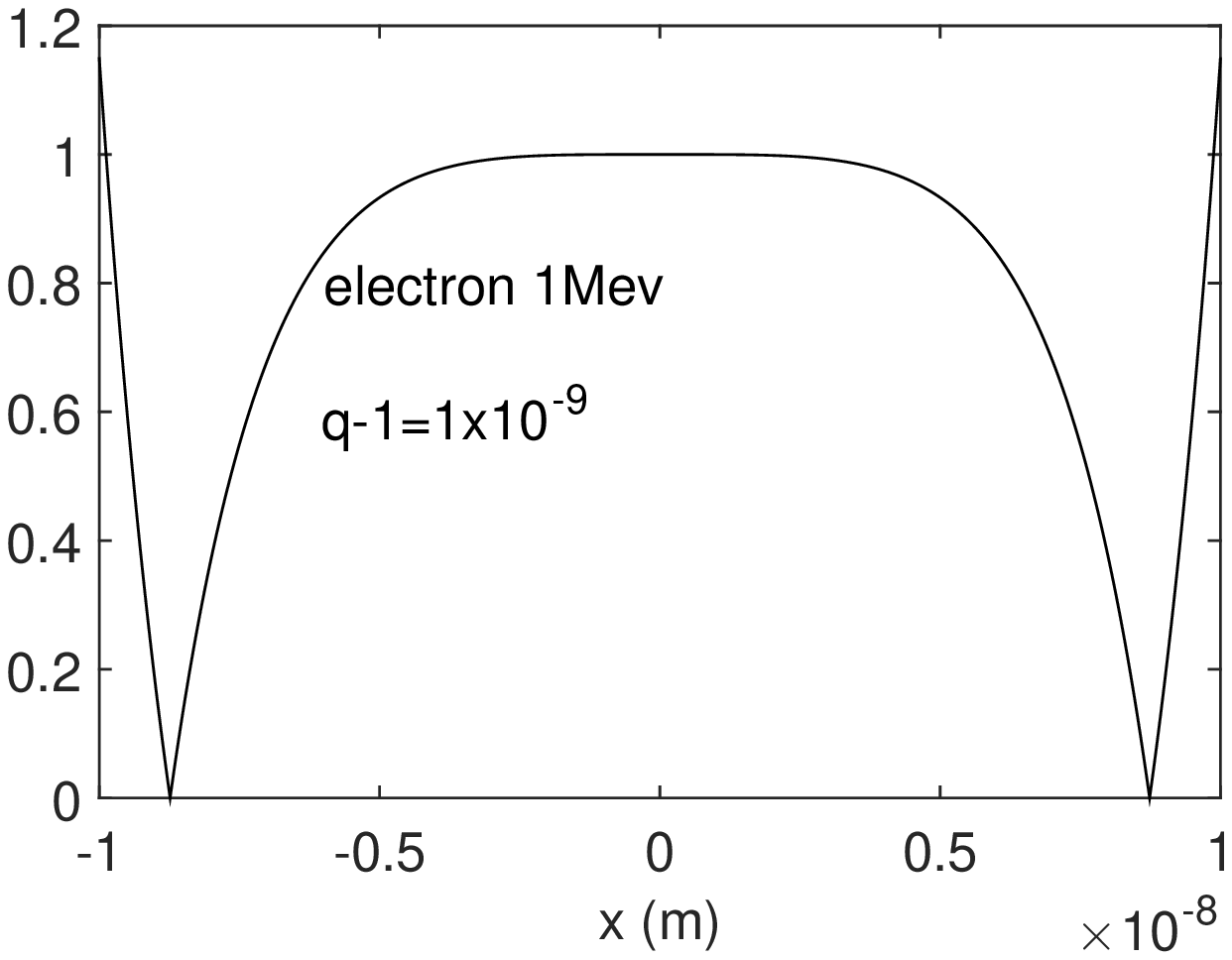}
\vspace{-0.2cm} \caption{Ratio R vs. $x$ (in meters) for 1 MeV electrons
and $q-1= 10^{-9}$.}
\end{center}
\end{figure}

\newpage

\begin{figure}[h]
\begin{center}
\includegraphics[scale=0.6,angle=0]{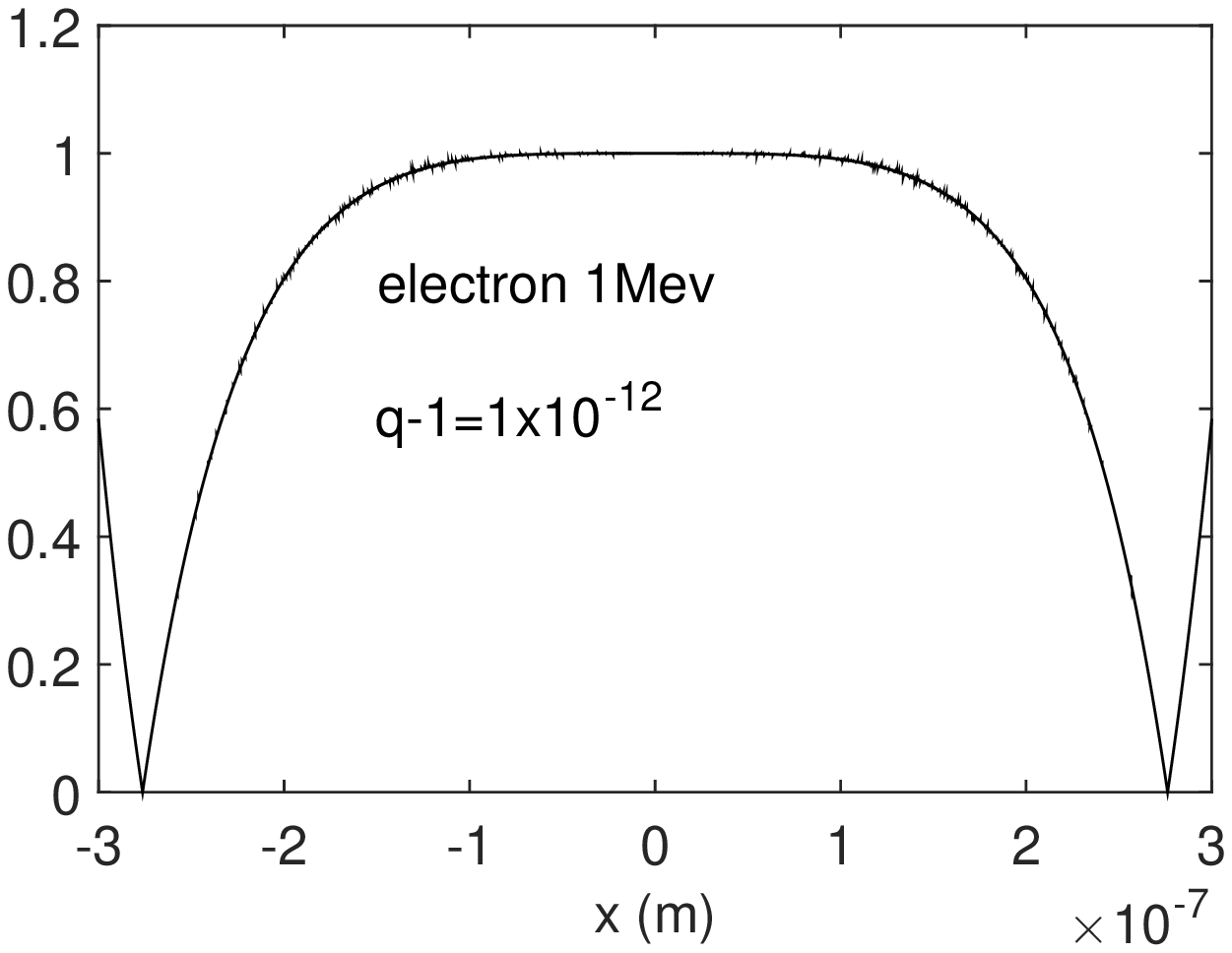}
\vspace{-0.2cm} \caption{Ratio R vs. $x$ (in meters) for 1 MeV electrons
and $q-1= 10^{-12}$.}
\end{center}
\end{figure}

\newpage

\begin{figure}[h]
\begin{center}
\includegraphics[scale=0.6,angle=0]{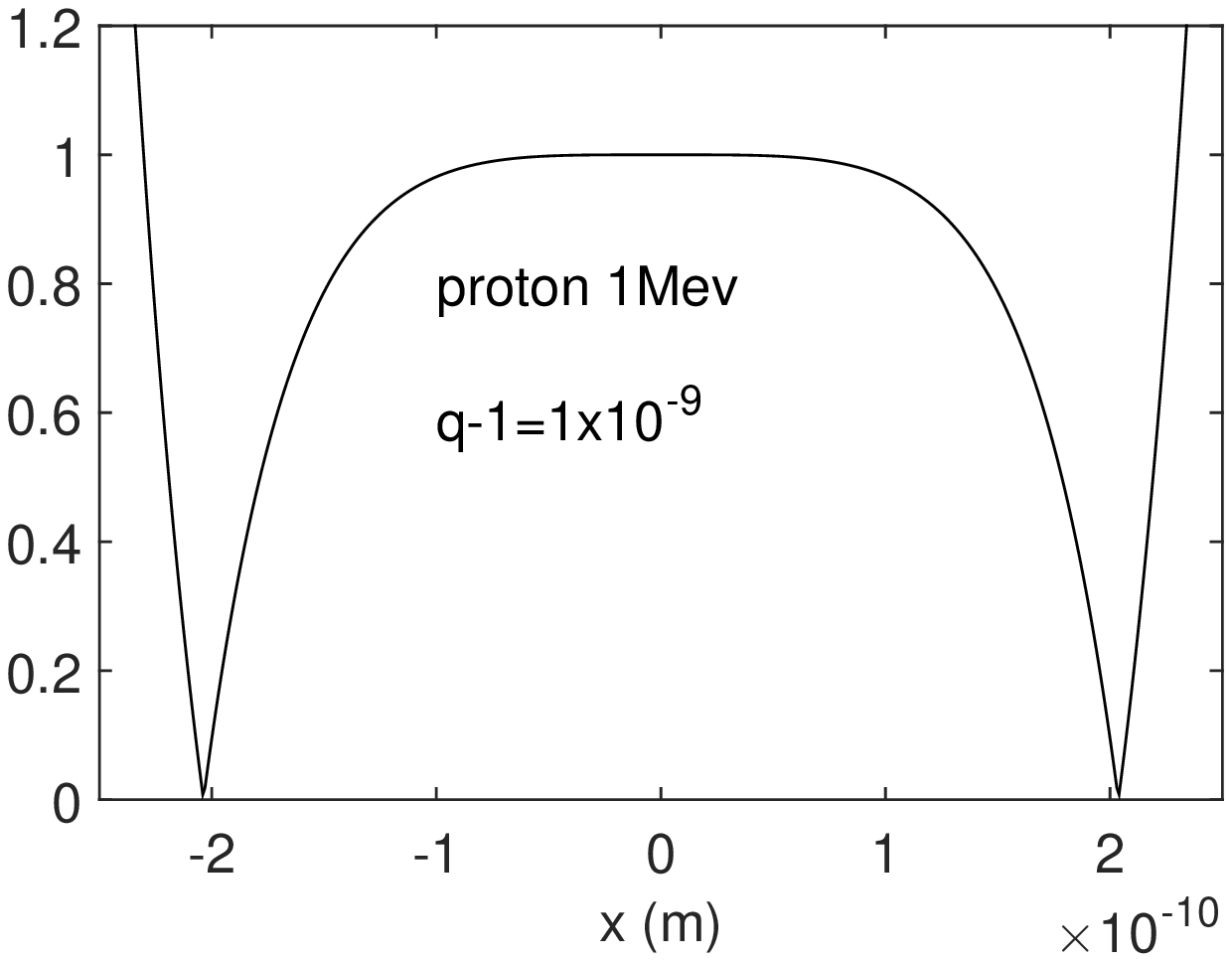}
\vspace{-0.2cm} \caption{Ratio R vs. $x$ (in meters) for 1 MeV protons
and $q-1= 10^{-9}$.}
\end{center}
\end{figure}

\newpage
\begin{figure}[h]
\begin{center}
\includegraphics[scale=0.6,angle=0]{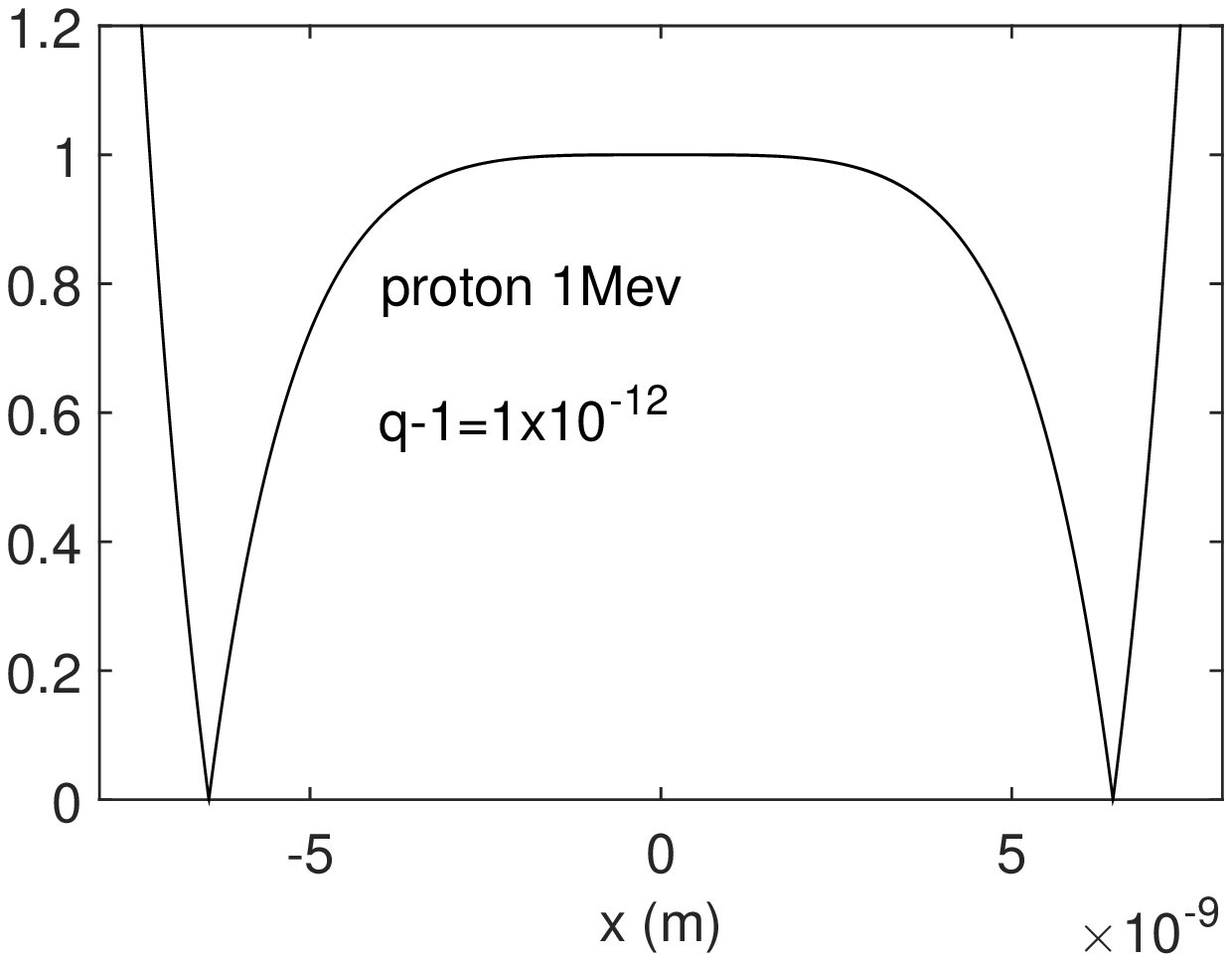}
\vspace{-0.2cm} \caption{Ratio R vs. $x$ (in meters) for 1 MeV protons
and $q-1= 10^{-12}$.}
\end{center}
\end{figure}

\newpage
\begin{figure}[h]
\begin{center}
\includegraphics[scale=0.6,angle=0]{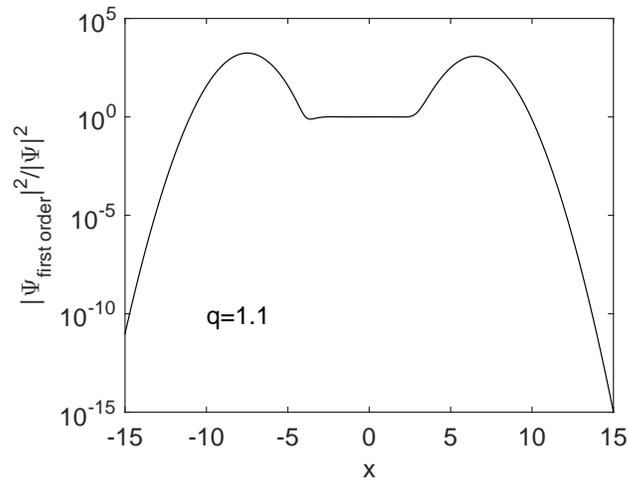}
\vspace{-0.2cm} \caption{q-Gaussian ratio of Eq. (4.12) over Eq. (4.1) vs. $x$ (in absolute units) for
$q-1= 10^{-3}$.}
\end{center}
\end{figure}

\newpage

\renewcommand{\thesection}{\Alph{section}}

\renewcommand{\theequation}{\Alph{section}.\arabic{equation}}

\setcounter{section}{1}

\section*{Appendix A}

\setcounter{equation}{0}

\subsection*{First order expansion of $\psi =e_{q}$}

 We  write
\begin{equation*}
\psi=e^{\frac{1}{1-q}\ln\corch},
\end{equation*}
so that
\begin{equation}
\label{eq1.7} \dq{\psi}=
\dq{\left\{\frac{1}{1-q}\ln\corch\right\}}\psi,
\end{equation}
\begin{equation*}
\dq{\psi}= \left\{\frac{1}{(1-q)^2}\ln\corch-\frac{z}{(1-q)\corch}
\right\}\psi.
\end{equation*}
As $ln(1+x)=x-\frac{x^2}{2}+\cdots$, we obtain
\begin{equation*}
\dq{\psi}=\left\{\frac{1}{(1-q)^2}\left[(1-q)z-\frac{(1-q)^{2}z^{2}}{2}\right]
-\frac{z}{(1-q)\corch}\right\}\psi,
\end{equation*}
or
\begin{equation}
\label{eq1.8}
\dq{\psi}=\left\{\frac{z}{(1-q)}-\frac{z}{(1-q)\corch}-\frac{z^2}{2}\
\right\}\psi.
\end{equation}
Let us  evaluate $\dq{\psi}$ at q=1:
\begin{equation*}
\dq{\psi}=\left\{z\left\{\frac{1}{(1-q)}-\frac{1}{(1-q)\corch}\right\}-\frac{z^2}{2}\right\}\psi,
\end{equation*}
\begin{equation*}
\dq{\psi}=\left\{z\frac{\corch-1}{(1-q)\corch}-\frac{z^2}{2}\right\}\psi.
\end{equation*}
When $q\to1$
\begin{equation*}
\frac{\corch-1}{(1-q)\corch}\to \frac{0}{0}.
\end{equation*}
Using now L'Hopital's rule one has
\begin{equation*}
\limq{\dq{\psi}}=\limq{\left\{\frac{z^2}
{\corch+(1-q)z}-\frac{z^2}{2}\right\}\psi},
\end{equation*}
\begin{equation*}
\val{\dq{\psi}}=\left(z^2-\frac{z^2}{2}\right)e^z,
\end{equation*}
\begin{equation}
\val{\dq{\psi}}=\frac{z^2}{2}e^z \label{eq1.9}.
\end{equation}
Thus, the first order Taylor's expansion of $\psi =e_{q}$ is
\begin{equation*}
\psi\simeq\val{\psi}+(q-1)\val{\dq{\psi}}.
\end{equation*}

\subsection*{Second derivative with respect to $x$}

The first order derivative respect to de variable x of $e_{q}$ is
\[\dx{\psi}=\frac{ip}{\hbar}e^{\z}-\frac{(q-1)}{2\hbar^2}\left[2p(px-Et)e^{\z}\right.\]
\begin{equation}
\label{eq1.12} \left.+\frac{ip}{\hbar}(px-Et)^2e^{\z}\right],
\end{equation}
For the second derivative we have
\[\dxs{\psi}=-\frac{p^2}{\hbar^2}e^{\z}-\frac{(q-1)}{2\hbar^2}\left[2p^2e^{\z}+\frac{4ip^2}{\hbar}(px-Et)e^{\z}\right.\]
\[\left.-\frac{p^2}{\hbar^2}(px-Et)^2e^{\z}\right],\]
or
\[\dxs{\psi}=-\frac{p^2}{\hbar^2}e^{\z}-(q-1)\frac{p^2}
{\hbar^2}e^{\z}-\]
\[(q-1)\frac{2ip^2}{\hbar^3}(px-Et)e^{\z}
+(q-1)\frac{p^2}{2\hbar^4}(px-Et)^2e^{\z}\].
\setcounter{section}{2}

\section*{Appendix B}

\setcounter{equation}{0}

\subsection*{First order expansion of f(t)}
Since  we can write
\begin{equation*}
f(t)=\corchf^{\frac{1}{q-1}}=e^{\frac{1}{q-1}ln\corchf},
\end{equation*}
one has
\begin{equation*}
\dq{f}=\dq{(\frac{1}{q-1}\ln\corchf)}f
\end{equation*}
\[=\left\{-\frac{1}{(q-1)^2}ln\corchf+\right.\]
\[\left.\frac{1}{(q-1)\corchf}
\left[\frac{-iEt}{q\hbar}-\frac{(1-q)iEt}
{q^{2}\hbar}\right]\right\}f.\] Since
$\ln(1+x)=x-\frac{x^2}{2}+\cdots$, we have
\[\dq{f}=\left\{-\frac{1}{(q-1)^2}
\left[\frac{i}{\hbar}\frac{(1-q)}{q}Et+\frac{(1-q)^2}{2q^2}
\frac{i}{\hbar}Et\right]\right.\]
\begin{equation}
\label{eq2.5}
\left.-\frac{1}{(q-1)\corchf}\left[\frac{iEt}{q\hbar}+\frac{(1-q)}{q^2}\frac{iEt}{\hbar}\right]\right\}f.
\end{equation}
So as to evaluate $f$ at $q=1$ we rearrange terms in (\ref{eq2.5})
and obtain

\[\dq{f}=\left\{\frac{iEt}{q\hbar}
\left[-\frac{1}{(1-q)}+ \frac{1}{(1-q)\corchf}+\right.\right.\]
\begin{equation*}
\left.\left.\frac{1}{q\corchf}\right]-\frac{E^{2}t^2}{\hbar^2}
\frac{1}{2q^{2}}\right\}f.
\end{equation*}
Moreover,
\begin{equation}
\label{eq2.6} \dq{f}=\left\{\frac{iEt}{q\hbar}
\left[\frac{1-\corchf}{(1-q)\corchf}+\frac{1}{q\corchf}\right]
-\frac{E^{2}t^2}{\hbar^2}\frac{1}{2q^{2}}\right\}f.
\end{equation}
As $q\to1$
\begin{equation*}
\frac{1-\corchf}{(1-q)\corchf}\to \frac{0}{0}.
\end{equation*}
Using again  L'Hopital's rule we find
\[\limq{\frac{1-\corchf}{(1-q)\corchf}}=\]
\[\limq{\frac{\frac{iEt}{q\hbar}+\frac{iEt(1-q)}
{\hbar q^2}}{-\corchf+(1-q)[\frac{iEt}{q\hbar}+\frac{iEt(1-q)}{\hbar
q^2}}}=-\frac{iEt}{\hbar}.\] Thus,
\begin{equation*}
\val{\dq{f}}=\left\{\frac{iEt}{\hbar}\left[1-\frac{iEt}{\hbar}\right]-\frac{E^{2}t^2}{2\hbar^2}\right\}e^{\frac{-iEt}{\hbar}},
\end{equation*}
or
\begin{equation*}
\val{\dq{f}}=\left(\frac{iEt}{\hbar}+\frac{E^{2}t^2}{\hbar^2}-\frac{E^{2}t^2}{2\hbar^2}\right)e^{\frac{-iEt}{\hbar}}.
\end{equation*}
Moreover,
\begin{equation}
\label{eq2.7}
\val{\dq{f}}=\left(\frac{iEt}{\hbar}+\frac{E^{2}t^2}{2\hbar^2}\right)e^{\frac{-iEt}{\hbar}}.
\end{equation}
The first order Taylor's expansion of $f(t)$ is
\begin{equation*}
f(t)\simeq \val{f}+(q-1)\val{\dq{f}}.
\end{equation*}
Replacing here (\ref{eq2.7}) we have
\begin{equation*}
f(t)\simeq e^{\frac{-iEt}{\hbar}}+(q-1)
\left(\frac{iEt}{\hbar}+\frac{E^{2}t^2}{2\hbar^2}\right)
e^{\frac{-iEt}{\hbar}},
\end{equation*}
or
\begin{equation}
\label{eq2.8} f(t)\simeq
e^{\frac{-iEt}{\hbar}}\left[1+(q-1)\left(\frac{iEt}{\hbar}+\frac{E^{2}t^2}{2\hbar^2}\right)\right].
\end{equation}

\subsection*{First order expansion of $f^q$}

Following a similar procedure used above to obtain (\ref{eq1.16}) we
have
\begin{equation}
\label{eq2.9} \val{\dq{f^q}}=\val{f}\val{\ln f}+\val{\dq{f}}.
\end{equation}
We know that
\begin{itemize}
\item $\val{f}=e^{\frac{-iEt}{\hbar}}$
\item $\val{(\ln f)}=-\frac{iEt}{\hbar}$
\item $\val{\dq{f}}=\left(\frac{iEt}{\hbar}+\frac{E^{2}t^2}{2\hbar^2}\right)e^{\frac{-iEt}{\hbar}}.$
\end{itemize}
Thus,
\begin{equation*}
\val{\dq{f^q}}=-\frac{iEt}{\hbar}
e^{\frac{-iEt}{\hbar}}+e^{\frac{-iEt}{\hbar}}\left(\frac{iEt}{\hbar}+\frac{E^{2}t^2}{2\hbar^2}\right),
\end{equation*}
or
\begin{equation*}
\val{\dq{f^q}}=-\frac{iEt}{\hbar}
e^{\frac{-iEt}{\hbar}}+\frac{iEt}{\hbar}e^{\frac{-iEt}{\hbar}}+\frac{E^{2}t^2}{2\hbar^2}e^{\frac{-iEt}{\hbar}}.
\end{equation*}
Moreover,
\begin{equation}
\label{eq2.10}
\val{\dq{f^q}}=\frac{E^{2}t^2}{2\hbar^2}e^{\frac{-iEt}{\hbar}}.
\end{equation}
The first order expansion of $f^q$ is
\begin{equation*}
f^q\simeq \val{f^q}+(q-1)\val{\dq{f^q}}.
\end{equation*}
Replacing here equation (\ref{eq2.10}) we find
\begin{equation*}
f^q\simeq
e^{\frac{-iEt}{\hbar}}+(q-1)\frac{E^{2}t^2}{2\hbar^2}e^{\frac{-iEt}{\hbar}},
\end{equation*}
or, finally,
\begin{equation}
\label{eq2.11} f^q\simeq e^{\frac{-iEt}{\hbar}}\left[1+(q-1)
\frac{E^{2}t^2}{2\hbar^2}\right].
\end{equation}

\subsection*{Derivative of $f^q$ with respect to t}

The first order derivative with respect to $t$ of $f^q$ is
\begin{equation*}
\dt{f^q}=-\frac{iE}{\hbar}e^{\frac{-iEt}{\hbar}}\left[1+(q-1)\frac{E^{2}t^2}{2\hbar^2}\right]+e^{\frac{-iEt}{\hbar}}\frac{(q-1)E^{2}t}{\hbar^2},
\end{equation*}
or
\begin{equation}
\label{eq2.12}
\dt{f^q}=-\frac{iE}{\hbar}e^{\frac{-iEt}{\hbar}}\left[1+(q-1)\frac{E^{2}t^2}{2\hbar^2}+(q-1)\frac{iEt}{\hbar}\right].
\end{equation}
\setcounter{section}{3}

\section*{Appendix C}

\setcounter{equation}{0}

\subsection*{Expansion of g(x)}

We can write
\begin{equation*}
g(x)=\left[1+\frac{i}{\hbar}\frac{(1-q)}{\sqrt{2(q+1)}}
px\right]^{\frac{2}{1-q}}=e^{\frac{2}{1-q}\ln\corchg},
\end{equation*}
so that:
\begin{equation*}
\dq{g}=\dq{\left\{\frac{2}{1-q}\ln\corchg\right\}}g,
\end{equation*}
\[=\left\{\frac{2}{(1-q)^2}\ln\corchg+\right.\]
\[\frac{2}{(1-q)\corchg}
\left[\frac{-ipx}{\hbar}\frac{1}{\sqrt{2(q+1)}}\right.\]
\[\left.\left.-\frac{ipx}{\hbar}\frac{(1-q)}{[2(q+1)]^{3/2}}
\right]\right\}g.\] Since  $\ln(1+x)=x-\frac{x^2}{2}+\cdots$, we
have
\begin{multline}
\label{eq2.13} \dq{g}=\left\{\frac{2}{(1-q)^2}\left[
\frac{ipx}{\hbar}\frac{(1-q)}{\sqrt{2(q+1)}}+\frac{(1-q)^2}{2(q+1)}\frac{p^{2}x^2}{2\hbar^2}\right]\right.\\
\left.+\frac{2}{(1-q)\corchg}
\left[-\frac{ipx}{\hbar}\frac{1}{\sqrt{2(q+1)}}-\frac{(1-q)}{[2(q+1)]^{3/2}}\frac{ipx}{\hbar}\right]\right\}g.
\end{multline}
So as to evaluate $g$ at $q=1$ we rearrange terms to obtain
\[\dq{g}=\left\{\frac{ipx}{\hbar}\frac{2}{\sqrt{2(q+1)}(1-q)}+\frac{p^{2}x^2}{\hbar^2}\frac{1}{2(q+1)}-\right.\]
\[\frac{ipx}{\hbar}\frac{2}{\sqrt{2(q+1)}(1-q)\corchg}\]
\[\left.-\frac{ipx}{\hbar}\frac{2}{[2(q+1)]^{3/2}\corchg}\right\}g.\]

Moreover,
\[\dq{g}=\left\{\frac{2ipx}{\hbar}\left[
\frac{\corchg-1}{(1-q)\sqrt{2(q+1)}\corchg}\right.\right.-\]
\[\left.\left.\frac{1}{[2(q+1)]^{3/2}\corchg}\right]
+\frac{p^{2}x^2}{\hbar^2}\frac{1}{2(q+1)}\right\}g.\] As $q\to1$,
\begin{equation}
\label{eq2.14} \frac{\corchg-1}{(1-q)\sqrt{2(q+1)}\corchg}\to
\frac{0}{0}.
\end{equation}
Appealing once again to L'Hopital's rule we find
\[\limq{\frac{\corchg-1}{(1-q)\sqrt{2(q+1)}\corchg}}=\]
\[\limq{\frac{\frac{-ipx}{\hbar}\frac{1}{\sqrt{2(q+1)}}-\frac{ipx}{\hbar}\frac{(1-q)}{[2(q+1)]^{3/2}}}{-\sqrt{2(q+1)}\corchg+(1-q)X}},\]
\begin{equation*}
=\frac{ipx}{4\hbar},
\end{equation*}
where
\[X=\frac{\corchg}{\sqrt{2(q+1)}}+\sqrt{2(q+1)}\left[\frac{-ipx}{\hbar\sqrt{2(q+1)}}-\frac{(1-q)ipx}{\hbar[2(q+1)]^{3/2}}\right].\]
Thus,
\begin{equation*}
\val{\dq{g}}=\left\{\frac{2ipx}{\hbar}\left[\frac{ipx}{4\hbar}-\frac{1}{8}\right]+\frac{p^{2}x^2}{4\hbar^2}\right\}e^{\frac{ipx}{\hbar}},
\end{equation*}
or
\begin{equation*}
\val{\dq{g}}=\left(-\frac{ipx}{4\hbar}-\frac{p^{2}x^2}{2\hbar^2}+\frac{p^{2}x^2}{4\hbar^2}\right)e^{\frac{ipx}{\hbar}},
\end{equation*}
and, finally,
\begin{equation}
\label{eq2.15}
\val{\dq{g}}=-\frac{1}{4}\left(\frac{ipx}{\hbar}+\frac{p^{2}x^2}{\hbar^2}\right)e^{\frac{ipx}{\hbar}}.
\end{equation}
The first order expansion of $g(x)$ is
\begin{equation*}
g(x)\simeq \val{g}+(q-1)\val{\dq{g}}.
\end{equation*}
Replacing here (\ref{eq2.15}) we get
\begin{equation*}
g(x)\simeq
e^{\frac{ipx}{\hbar}}+\frac{(1-q)}{4}\left(\frac{ipx}{\hbar}+\frac{p^{2}x^2}{\hbar^2}\right)e^{\frac{ipx}{\hbar}},
\end{equation*}
or,
\begin{equation}
\label{eq2.16} g(x)\simeq
e^{\frac{ipx}{\hbar}}\left[1+\frac{(1-q)}{4}\left(\frac{ipx}{\hbar}+\frac{p^{2}x^2}{\hbar^2}\right)\right].
\end{equation}

\subsection*{Second derivative of g with respect to x}

The first order derivative of $g$ is
\begin{equation}
\label{eq2.17}
\dx{g}=\frac{ip}{\hbar}e^{\frac{ipx}{\hbar}}\left[1+\frac{(1-q)}{4}\left(\frac{ipx}{\hbar}+\frac{p^{2}x^2}{\hbar^2}\right)\right]+e^{\frac{ipx}{\hbar}}\frac{(1-q)}{4}\left(\frac{ip}{\hbar}+\frac{2xp^2}{\hbar^2}\right).
\end{equation}
For the second order derivative we have
\[\dxs{g}=\frac{-p^2}{\hbar^2}e^{\frac{ipx}{\hbar}}\left[1+\frac{(1-q)}{4}\left(\frac{ipx}{\hbar}+\frac{p^{2}x^2}{\hbar^2}\right)\right]+\frac{ip}{\hbar}e^{\frac{ipx}{\hbar}}\frac{(1-q)}{4}\left(\frac{ip}{\hbar}+\frac{2xp^2}{\hbar^2}\right)\]
\[+\frac{ip}{\hbar}e^{\frac{ipx}{\hbar}}\frac{(1-q)}{4}\left(\frac{ip}{\hbar}+\frac{2xp^2}{\hbar^2}\right)+e^{\frac{ipx}{\hbar}}\frac{(1-q)}{4}\frac{2p^2}{\hbar^2},\]
or
\begin{equation*}
\dxs{g}=-\frac{p^2}{\hbar^2}e^{\frac{ipx}{\hbar}}\left[1+\frac{(1-q)}{4}\frac{ipx}{\hbar}+\frac{(1-q)}{4}\frac{p^{2}x^2}{\hbar^2}-(1-q)\frac{ipx}{\hbar}\right],
\end{equation*}
so that,  finally,
\begin{equation}
\label{eq2.18}
\dxs{g}=-\frac{p^2}{\hbar^2}e^{\frac{ipx}{\hbar}}\left[1-\frac{3(1-q)}{4}\frac{ipx}{\hbar}+\frac{(1-q)}{4}\frac{p^{2}x^2}{\hbar^2}\right].
\end{equation}

\subsection*{First order expansion of $g^q$}

Following a similar procedure to that used to obtain [1.5]
 in the paper  we have now
\begin{equation}
\label{eq2.19} \val{\dq{g^q}}=\val{g}\val{\ln g}+\val{\dq{g}}.
\end{equation}
We know that
\begin{itemize}
\item $\val{g}=e^{\frac{ipx}{\hbar}}$
\item $\val{(lng)}=\frac{ipx}{\hbar}$
\item $\val{\dq{g}}=-\frac{1}{4}\left(\frac{ipx}{\hbar}+\frac{p^{2}x^2}{\hbar^2}\right)e^{\frac{ipx}{\hbar}}.$
\end{itemize}
Thus,
\begin{equation*}
\val{\dq{g^q}}=\frac{ipx}{\hbar}
e^{\frac{ipx}{\hbar}}-\frac{1}{4}e^{\frac{ipx}{\hbar}}\left(\frac{ipx}{\hbar}+\frac{p^{2}x^2}{\hbar^2}\right),
\end{equation*}
or,
\begin{equation*}
\val{\dq{g^q}}=\frac{ipx}{\hbar}
e^{\frac{-iEt}{\hbar}}\left(1-\frac{1}{4}+\frac{ipx}{4\hbar}\right).
\end{equation*}
Moreover:
\begin{equation}
\label{eq2.20} \val{\dq{g^q}}=\frac{ipx}{\hbar}
e^{\frac{ipx}{\hbar}}\left(\frac{3}{4}+\frac{ipx}{4\hbar}\right).
\end{equation}
The first order expansion  of $g^q$ is
\begin{equation}
\label{eq2.21} g^q\simeq \val{g^q}+(q-1)\val{\dq{g^q}}.
\end{equation}
Replacing Eq. (\ref{eq2.20}) in this expansion we obtain
\begin{equation*}
g^q\simeq
e^{\frac{ipx}{\hbar}}+(q-1)\frac{ipx}{\hbar}e^{\frac{ipx}{\hbar}}\left(\frac{3}{4}+\frac{ipx}{4\hbar}\right),
\end{equation*}
or
\begin{equation}
\label{eq2.22} g^q\simeq
e^{\frac{ipx}{\hbar}}\left[1+\frac{3(q-1)}{4}\frac{ipx}{\hbar}-\frac{(q-1)}{4}\frac{p^{2}x^2}{\hbar^2}\right].
\end{equation}
\setcounter{section}{4}

\section*{Appendix D}

\setcounter{equation}{0}

\subsection*{Second derivative with respect to t}
The first order derivative of $F$ is
\[\dt{F}=-i\omega e^{i(kx-\omega t)}
\left[1+\frac{(1-q)}{2}(kx-\omega t)^2\right]-\omega e^{i(kx-\omega
t)}(1-q)(kx-\omega t),\] or
\begin{equation}
\label{eq3.3} \dt{F}=-i\omega e^{i(kx-\omega t)}
\left[1-(1-q)i(kx-\omega t)+\frac{(1-q)}{2}(kx-\omega t)^2\right].
\end{equation}
For the second order derivative we have
\[\dts{F}=-\omega^2 e^{i(kx-\omega t)}
\left[1-(1-q)i(kx-\omega t)+\frac{(1-q)}{2}(kx-\omega t)^2\right]\]
\[-i\omega e^{i(kx-\omega t)}
\left[i\omega (1-q)-\omega (1-q)(kx-\omega t)\right],\] or
\[\dts{F}=-\omega^{2}e^{i(kx-\omega t)}\left[1-(1-q)i(kx-\omega t)+\right.\]
\[\left.\frac{(1-q)}{2}(kx-\omega t)^2-(1-q)-(1-q)i(kx-\omega t)\right],\]
so that,  finally,
\begin{multline}
\label{eq3.5} \dts{F}=-\omega^2e^{i(kx-\omega t)}
\left[q+2i(q-1)(kx-\omega t)-\frac{(q-1)}{2}(kx-\omega t)^2\right].
\end{multline}

\subsection*{First order expansion of $qF^{2q-1}$}
The derivative of $qF^{2q-1}$ with respect to $q$ is
\begin{equation}
\label{eq3.6} \dq{(qF^{2q-1})}=F^{2q-1}+q\dq{F^{2q-1}}.
\end{equation}
We can write
\begin{equation*}
F^{2q-1}=e^{(2q-1)lnF},
\end{equation*}
so that
\begin{equation*}
\dq{F^{2q-1}}=\left[2lnF+\frac{(2q-1)}{F}\dq{F}\right]F^{2q-1},
\end{equation*}
\begin{equation}
\label{eq3.7} \dq{F^{2q-1}}=2F^{2q-1}lnF+(2q-1)F^{2q-2}\dq{F}.
\end{equation}
For the derivative of $qF^{2q-1}$ we have
\begin{equation}
\label{eq3.8}
\dq{(qF^{2q-1})}=F^{2q-1}+2qF^{2q-1}lnF+q(2q-1)F^{2q-2}\dq{F}.
\end{equation}
At $q=1$ we obtain
\begin{equation*}
\val{\dq{(qF^{2q-1})}}=\val{F}+2\val{F}\val{lnF}+\val{\dq{F}}.
\end{equation*}
We know that
\begin{itemize}
\item $\val{F}=e^{i(kx-\omega t)}$
\item $\val{(lnF)}=i(kx-\omega t)$
\item $\val{\dq{F}}=-\frac{(kx-\omega t)^2}{2}e^{i(kx-\omega t)}.$
\end{itemize}
Thus,
\begin{equation*}
\val{\dq{(qF^{2q-1})}}=e^{i(kx-\omega t)}+2i(kx-\omega t)
e^{i(kx-\omega t)}-\frac{(kx-\omega t)^2}{2}e^{i(kx-\omega t)},
\end{equation*}
or
\begin{equation}
\label{eq3.9} \val{\dq{(qF^{2q-1})}}=e^{i(kx-\omega
t)}\left[1+2i(kx-\omega t)-\frac{(kx-\omega t)^2}{2}\right].
\end{equation}
The first order expansion of $qF^{2q-1}$ is
\begin{equation}
\label{eq3.10} qF^{2q-1}\simeq
\val{(qF^{2q-1})}+(q-1)\val{\dq{(qF^{2q-1})}}.
\end{equation}
Replacing Eq. (\ref{eq3.9}) in this expansion we obtain
\begin{equation*}
qF^{2q-1}\simeq e^{i(kx-\omega t)}+(q-1) e^{i(kx-\omega
t)}\left[1+2i(kx-\omega t)-\frac{(kx-\omega t)^2}{2}\right],
\end{equation*}
or
\begin{equation}
\label{eq3.11} qF^{2q-1}\simeq e^{i(kx-\omega
t)}\left[q+2i(q-1)(kx-\omega t)-(q-1)\frac{(kx-\omega
t)^2}{2}\right].
\end{equation}

\end{document}